\documentclass{emulateapj}
\usepackage{amsmath}
\usepackage{txfonts}
\usepackage{natbib}
\usepackage{graphicx}
\usepackage{color}
\usepackage{multirow}
\usepackage{array}

\newcommand{\ud}{\ensuremath{\mathrm{d}}}

\shorttitle{Neutrino-driven explosion of 20 solar mass progenitor in three dimensions}

\shortauthors{Melson et al.}

\begin{document}

\title{Neutrino-driven explosion of a 20 solar-mass star in three dimensions\\
enabled by strange-quark contributions to neutrino-nucleon scattering}

\author{Tobias Melson\altaffilmark{1,2}, Hans-Thomas Janka\altaffilmark{1}, 
Robert Bollig\altaffilmark{1,2}, Florian Hanke\altaffilmark{1,2},
Andreas Marek\altaffilmark{3}, and Bernhard M\"uller\altaffilmark{4}}

\altaffiltext{1}{Max-Planck-Institut f\"ur Astrophysik, Karl-Schwarzschild-Str.~1, 
85748 Garching, Germany}
\altaffiltext{2}{Physik Department, Technische Universit\"at M\"unchen, 
James-Franck-Str. 1, 85748 Garching, Germany}
\altaffiltext{3}{Rechenzentrum der Max-Planck-Gesellschaft (RZG), Gie\ss enbachstr.~2,
85748 Garching, Germany}
\altaffiltext{4}{Monash Centre for Astrophysics, School of Physics and Astronomy, 
9 Rainforest Walk, Monash University, VIC 3800, Australia}
%\email{melson@mpa-garching.mpg.de, thj@mpa-garching.mpg.de}

\begin{abstract}
Interactions with neutrons and protons play a crucial role for the neutrino opacity
of matter in the supernova core. Their current implementation in many simulation
codes, however, is rather schematic and ignores not only modifications for the 
correlated nuclear medium of the nascent neutron star, but also free-space corrections
from nucleon recoil, weak magnetism or strange quarks, which can easily add up to
changes of several 10\% for neutrino energies in the spectral peak. 
In the Garching supernova simulations with the \textsc{Prometheus-Vertex} code, such
sophistications have been included for a long time except for the strange-quark 
contributions to the nucleon spin, which affect neutral-current neutrino scattering. 
We demonstrate on the basis of a 20\,$M_\odot$ progenitor star that a moderate
strangeness-dependent contribution of $g_\mathrm{a}^\mathrm{s} = -0.2$ to the axial-vector
coupling constant $g_\mathrm{a} \approx 1.26$ can turn an unsuccessful three-dimensional
(3D) model into a successful explosion. Such a modification is in the direction of
current experimental results and reduces the neutral-current scattering opacity of 
neutrons, which dominate in the medium around and above the neutrinosphere. This leads
to increased luminosities and mean energies of all neutrino species and strengthens
the neutrino-energy deposition in the heating layer. Higher nonradial kinetic energy
in the gain layer signals enhanced buoyancy activity that enables the onset
of the explosion at $\sim$300\,ms after bounce, in contrast to the model with vanishing
strangeness contributions to neutrino-nucleon scattering. Our results demonstrate the
close proximity to explosion of the previously published, unsuccessful 3D models of
the Garching group.
\end{abstract}


\section{Introduction}

According to the standard paradigm of the explosion mechanism of 
core-collapse supernovae (SNe), neutrino heating above the gain radius
initiates the outward acceleration of the stalled bounce shock and
provides a major fraction of the energy that unbinds the explosion
ejecta and powers the SN blast wave
\citep[e.g.,][]{colgate66,bethe85,bethe90,janka12,burrows13}.
Except for stars near the low-mass end of SN progenitors with
oxygen-neon-magnesium cores or small iron cores and a steep density
gradient of the overlying shells, successful explosions cannot be
obtained in spherical symmetry (1D) by state-of-the-art simulations. 
For more massive progenitors the success of the neutrino-driven
mechanism requires the support by multi-dimensional hydrodynamic
flows in the postshock layer associated with convective overturn and
the standing accretion shock instability \citep[SASI;][]{blondin03}.
Such flows increase the neutrino-energy deposition and create
buoyancy and turbulent pressure, thus reducing the critical threshold 
for the neutrino luminosity to trigger the explosive runaway of the shock
\citep[e.g.,][]{herant94,burrows95,janka96,murphy08,nordhaus10,hanke12,dolence13,murphy13}.
Breaking spherical symmetry
also allows for simultaneous mass accretion and outflow from the
proto-neutron star (PNS) after shock revival, which steepens the rise of
the explosion energy \citep{marek09}.

The first self-consistent three-dimensional (3D) stellar core-collapse
simulations with successful explosions were performed by \citet{fryer02,fryer04}, 
using smoothed-particle hydrodynamics and
a simple, grey diffusion description of neutrino transport, which
favored rapid explosions in 2D and 3D. Recently, the Garching group has
obtained a 3D explosion for a 9.6\,$M_\odot$ star with sophisticated,
state-of-the-art ray-by-ray-plus (RbR+), multi-group transport and found 
that 3D hydrodynamic flows enhance the explosion energy compared to the
2D (axisymmetric) case \citep{melson15}.
However, previous 3D simulations of this group for
11.2, 20, and 27\,$M_\odot$ progenitors (covering $\sim$400--550\,ms
after bounce) did not produce explosions although corresponding
2D models exploded \citep{hanke13,tamborra13,tamborra14a,tamborra14b}.
This is consistent with studies based on parametrized neutrino heating
\citep[e.g.,][]{hanke12,couch13,couch14,couchott15}
and self-consistent simulations with approximate neutrino transport
\citep{takiwaki12,takiwaki14}, which revealed that the onset of
explosion is less favored or delayed in 3D compared to 2D, contradicting
claims by \citet{nordhaus10} and, though moderated, by \citet{burrows12}
and \citet{dolence13}. Also \citet{lentz15}, using high-fidelity RbR
multi-group neutrino
diffusion, highlight a successful 15\,$M_\odot$ explosion that sets in
significantly later than the corresponding 2D model. 
Large-scale deformation modes seem to have the strongest supportive 
effect on the explosion but are weakened by the forward turbulent energy
cascade to small-scale structures in 3D, which is opposite to the 
2D case.

\begin{figure*}[t!]
\begin{center}
\includegraphics[width=.33\textwidth]{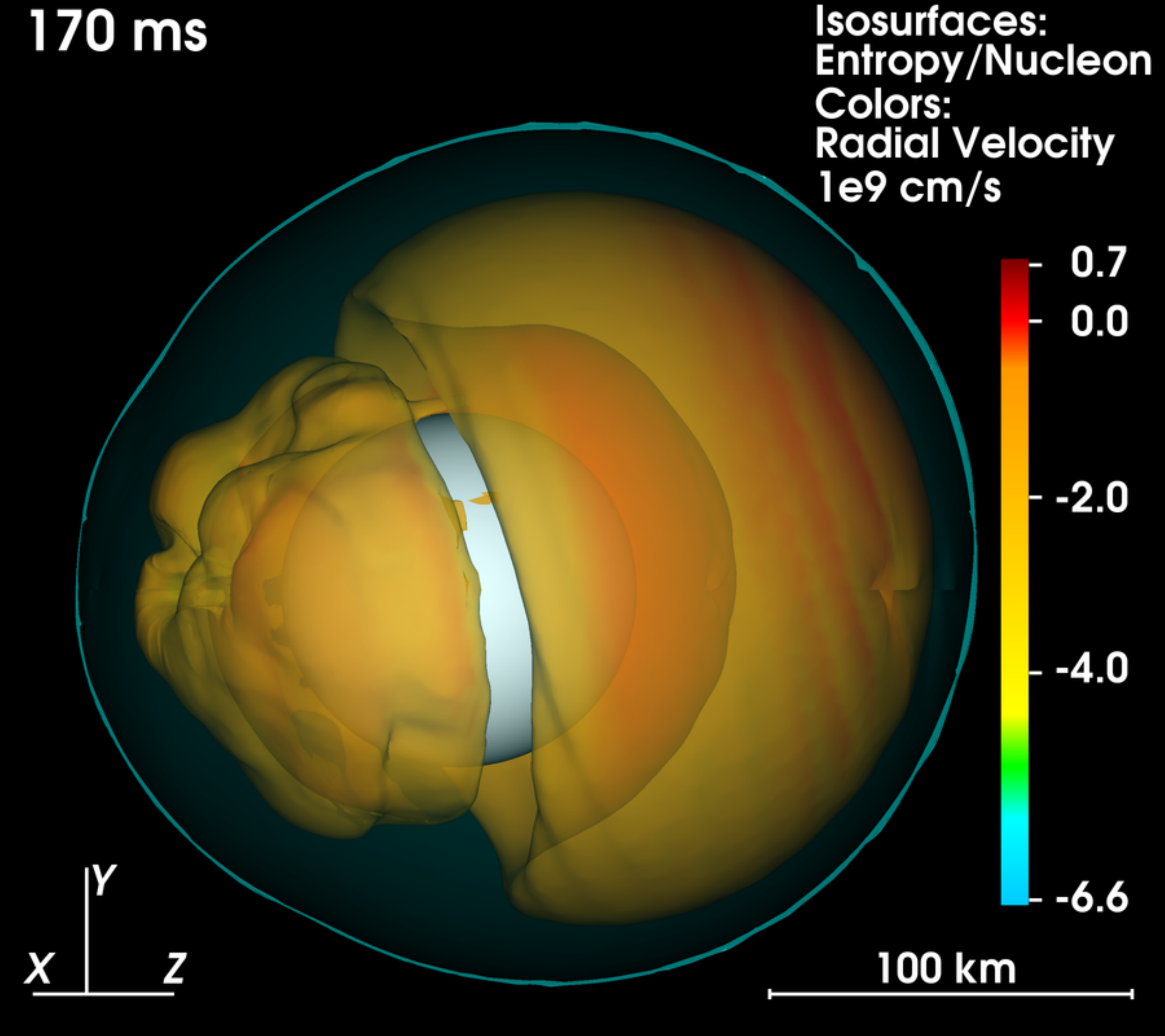}\hskip1.0pt
\includegraphics[width=.33\textwidth]{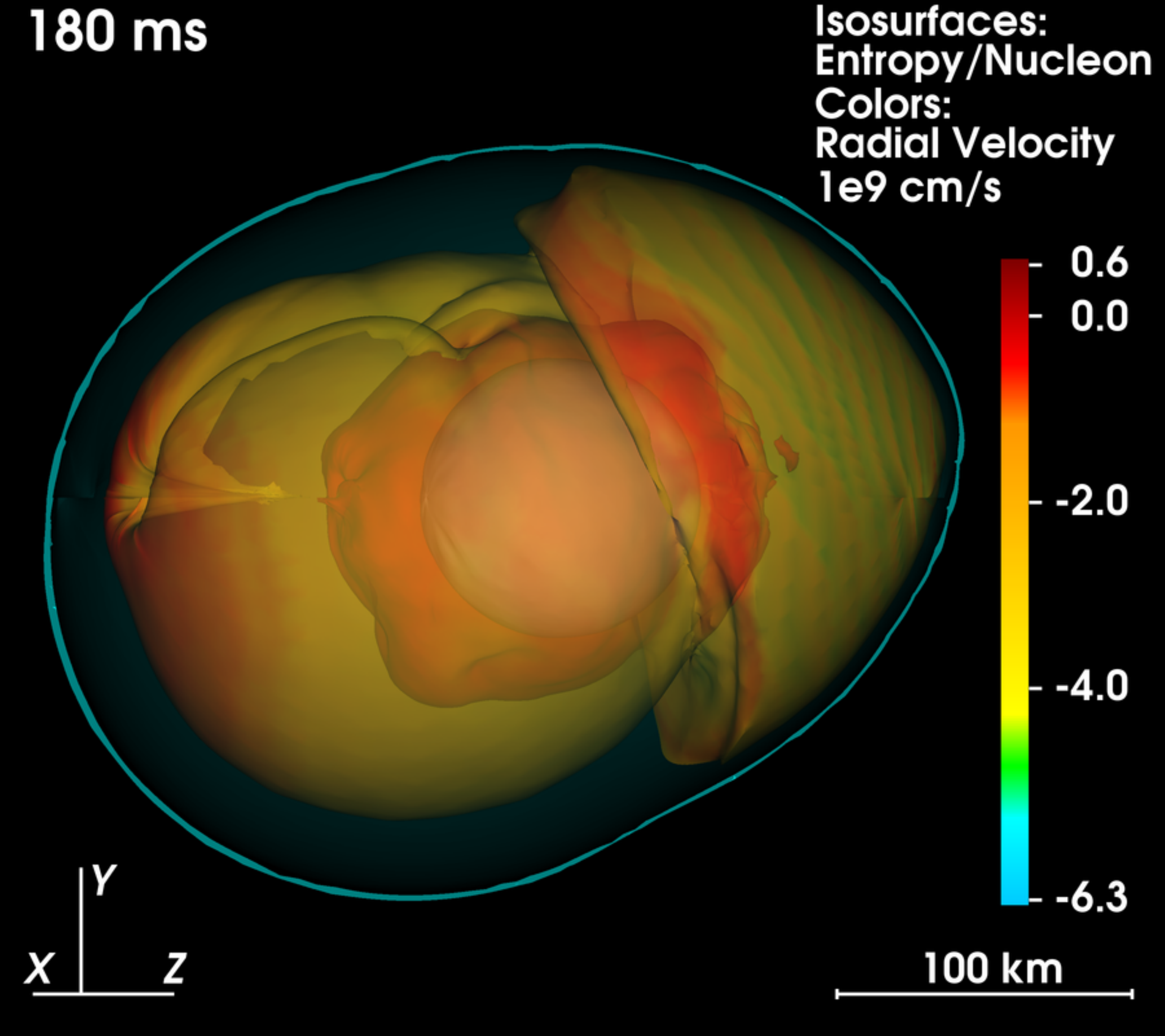}\hskip1.0pt
\includegraphics[width=.33\textwidth]{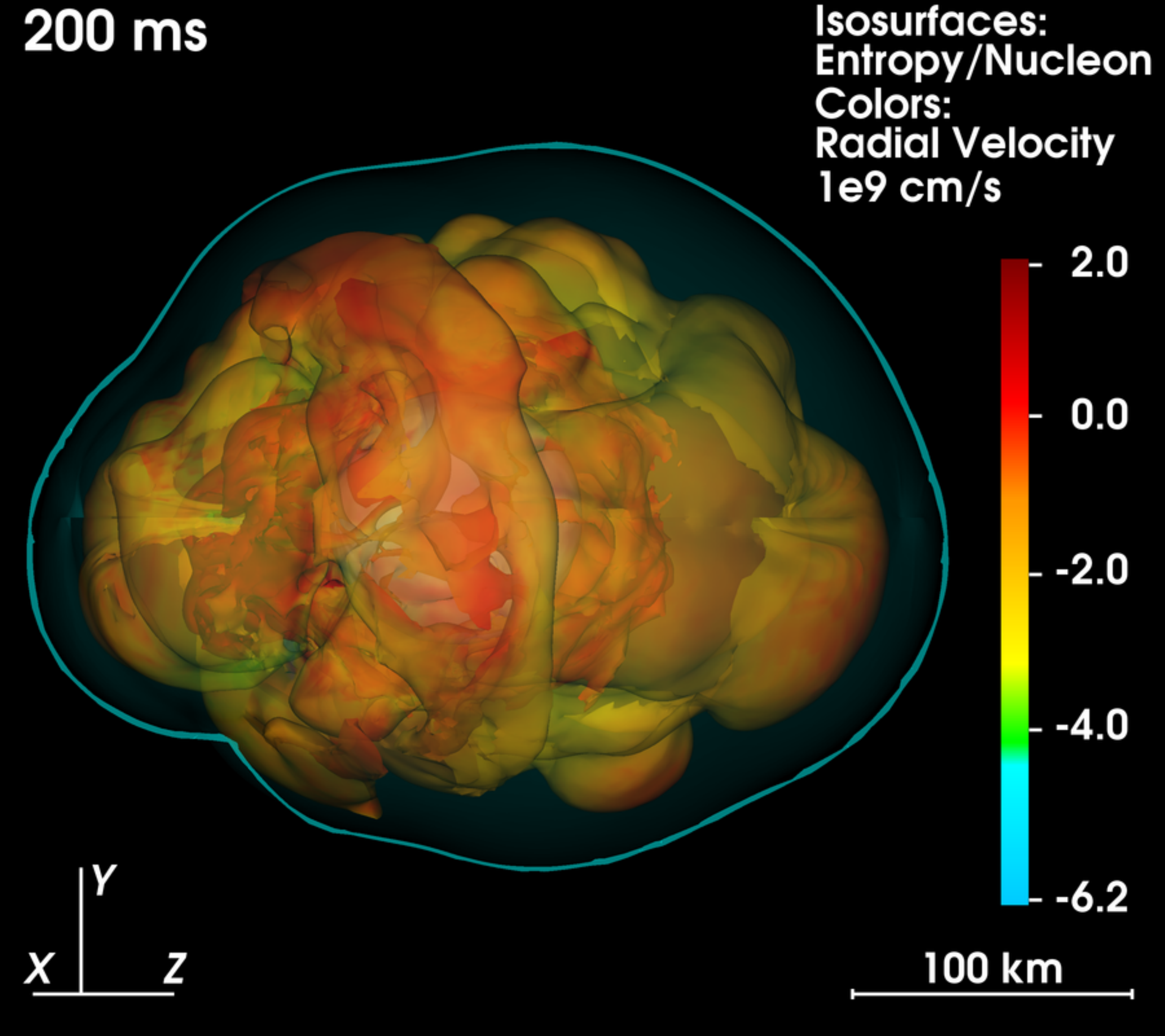}\\
\includegraphics[width=.33\textwidth]{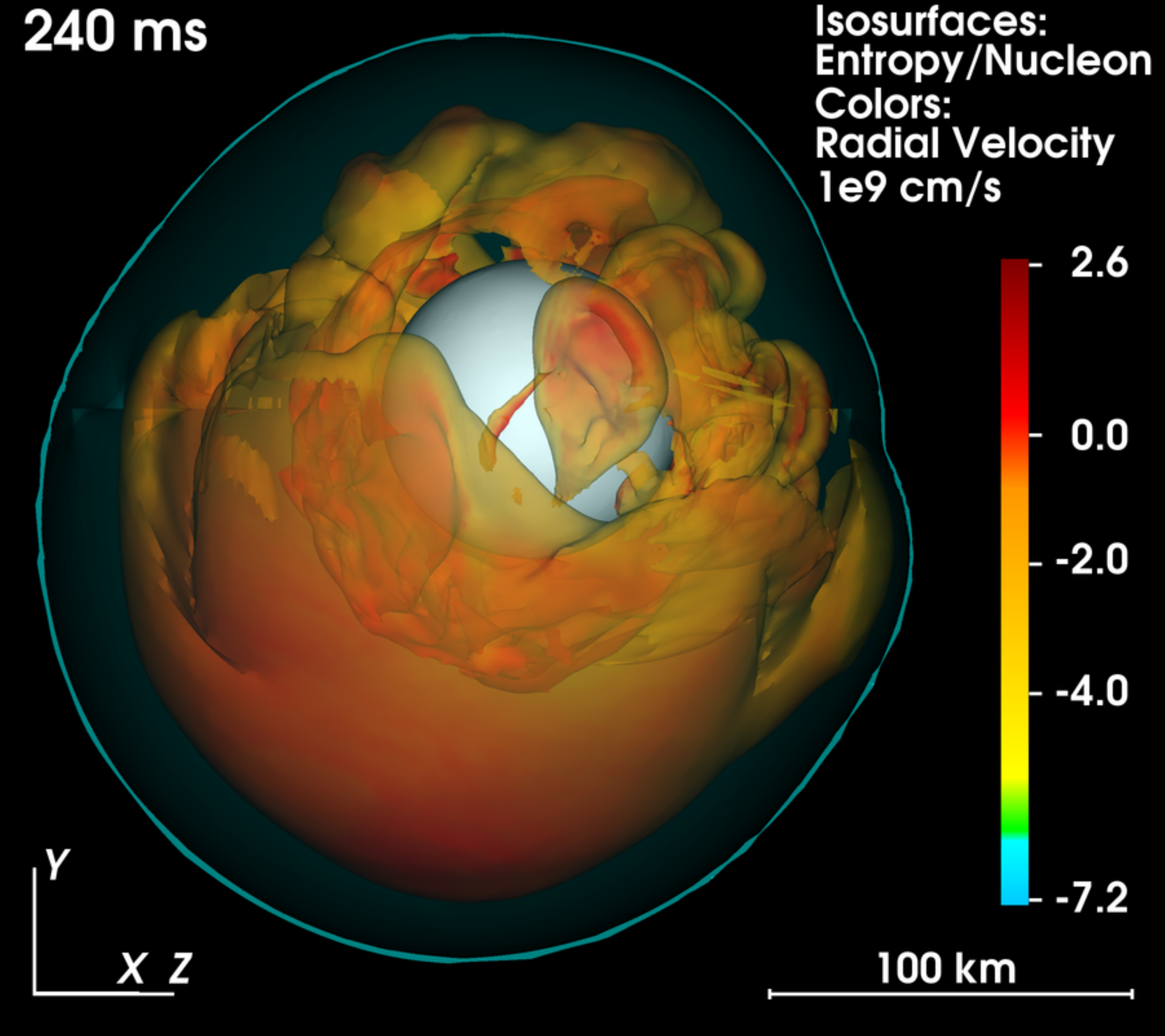}\hskip1.0pt
\includegraphics[width=.33\textwidth]{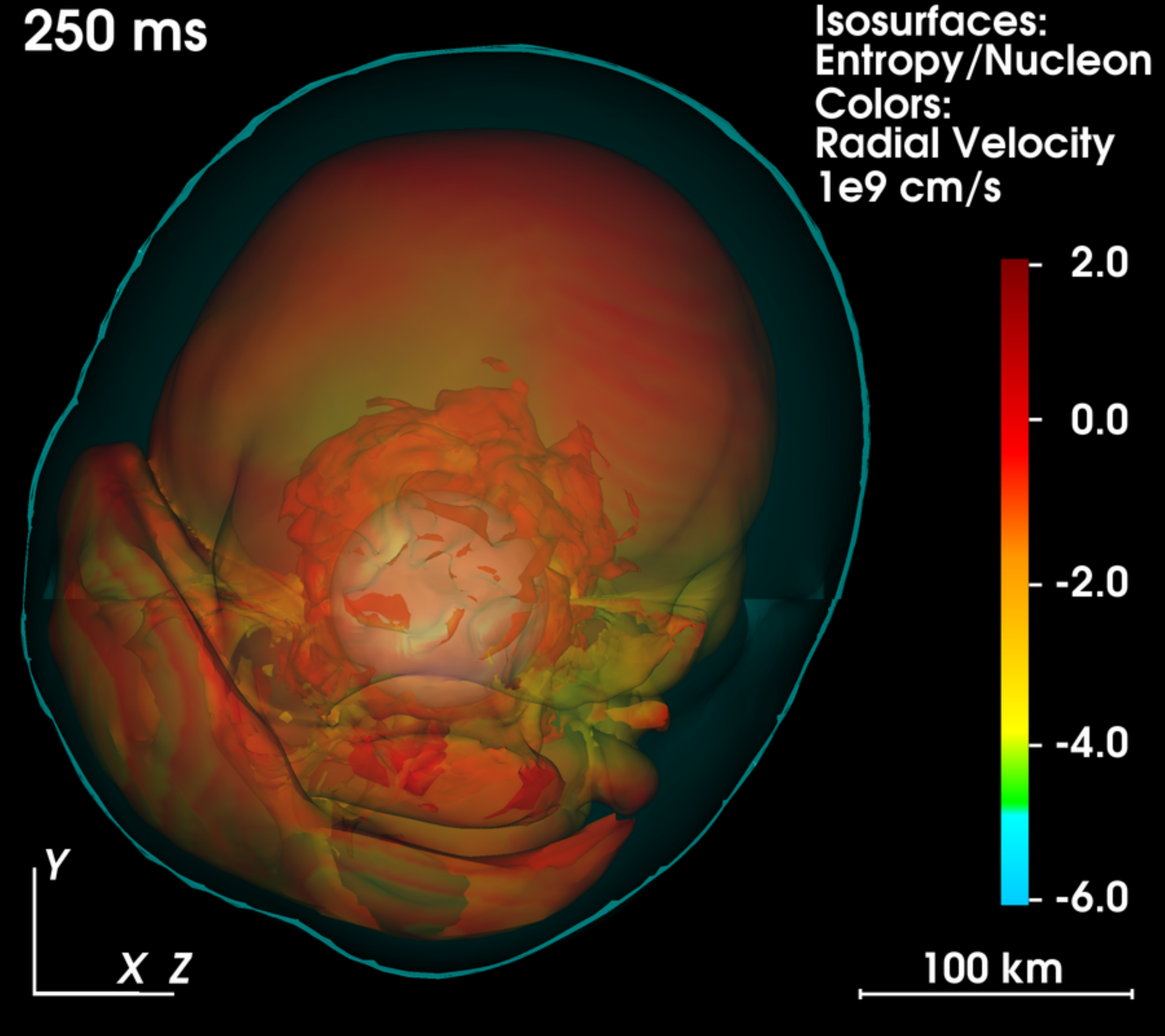}\hskip1.0pt
\includegraphics[width=.33\textwidth]{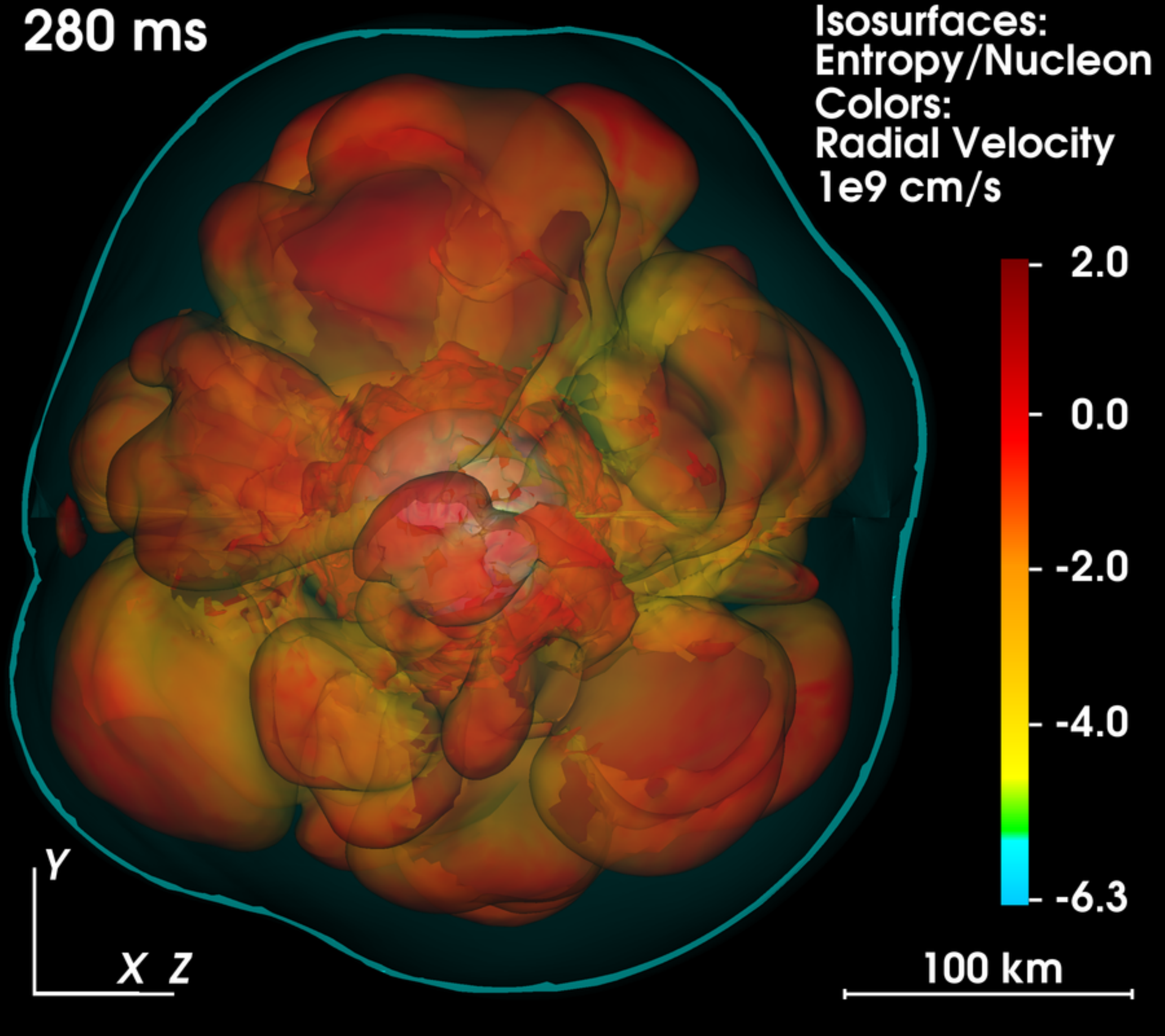}\\
\includegraphics[width=.33\textwidth]{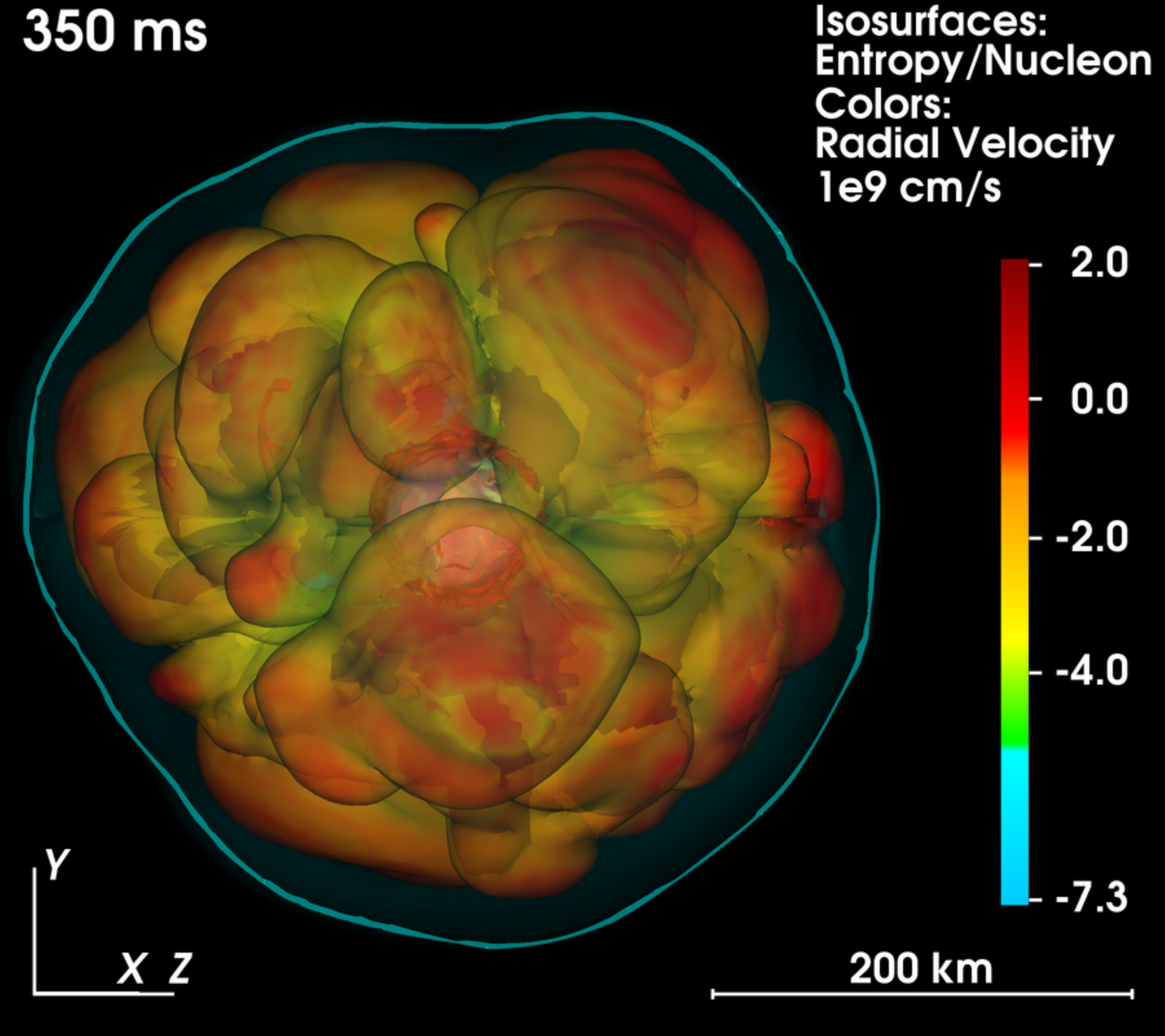}\hskip1.0pt
\includegraphics[width=.33\textwidth]{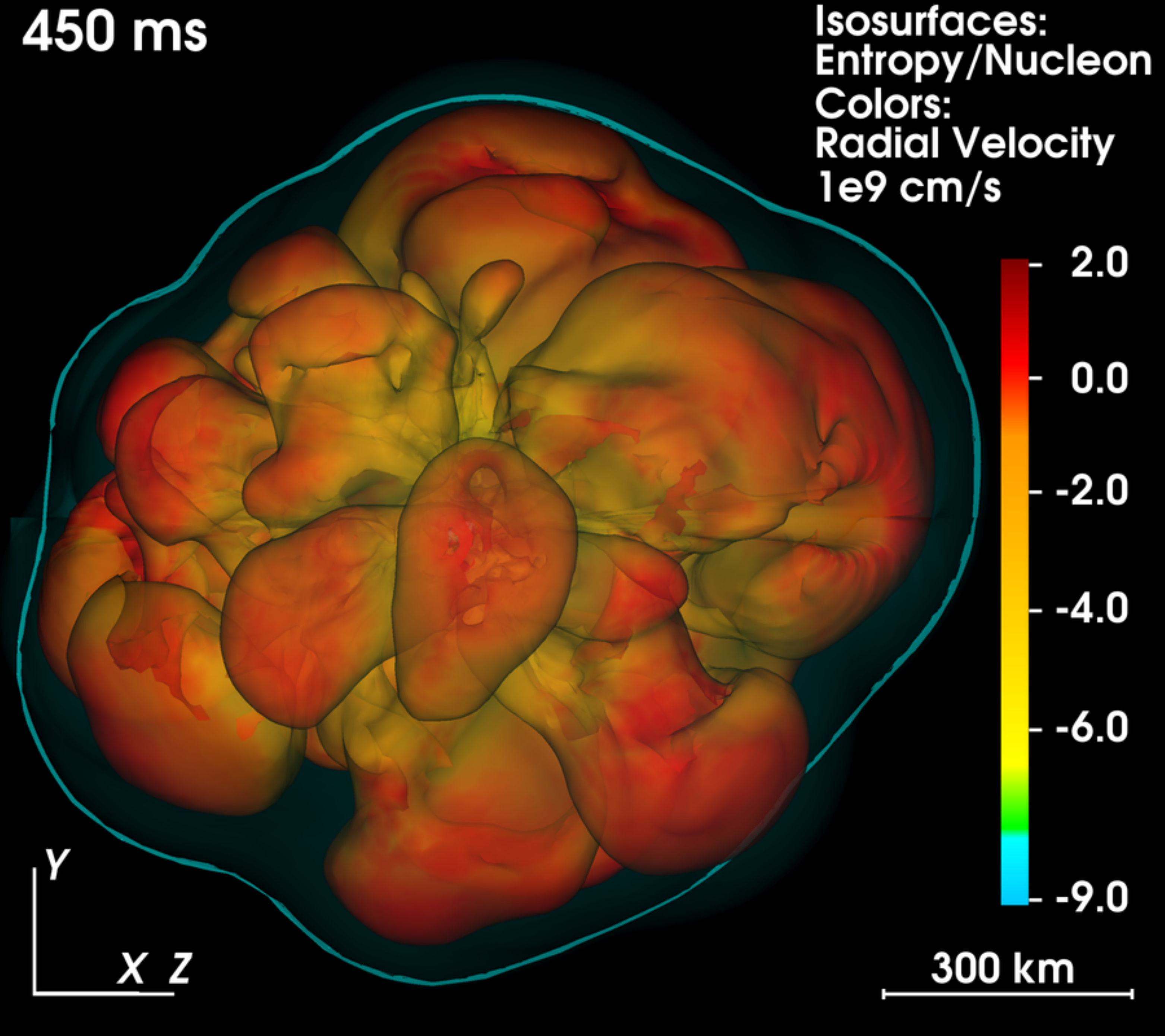}\hskip1.0pt
\includegraphics[width=.33\textwidth]{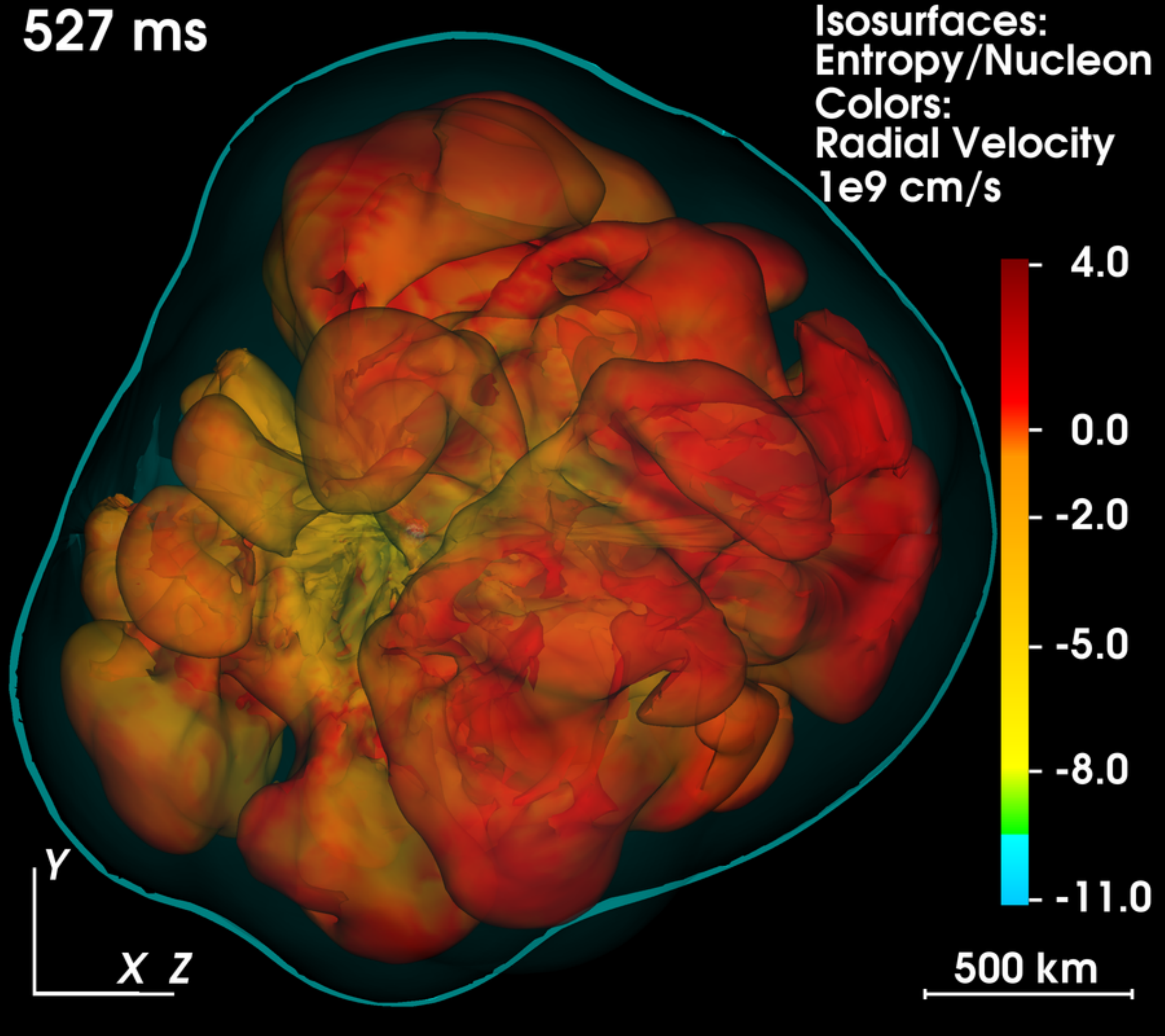}
\caption{
3D iso-entropy surfaces for different times after bounce for exploding
model 3Ds. Colors
represent radial velocities. The supernova shock is visible by a thin
surrounding line, the proto-neutron star by a whitish iso-density surface
of $10^{11}\,$g\,cm$^{-3}$. The yardstick indicates the length scale.
Strong SASI activity occurs between $\sim$120\,ms and $\sim$280\,ms.
{\em (An animation and interactive version of this figure are available in
the HTML version of this article.)}
}
\label{fig:3Dsevol}
\end{center}
\end{figure*}

\begin{figure*}
\includegraphics[width=\textwidth]{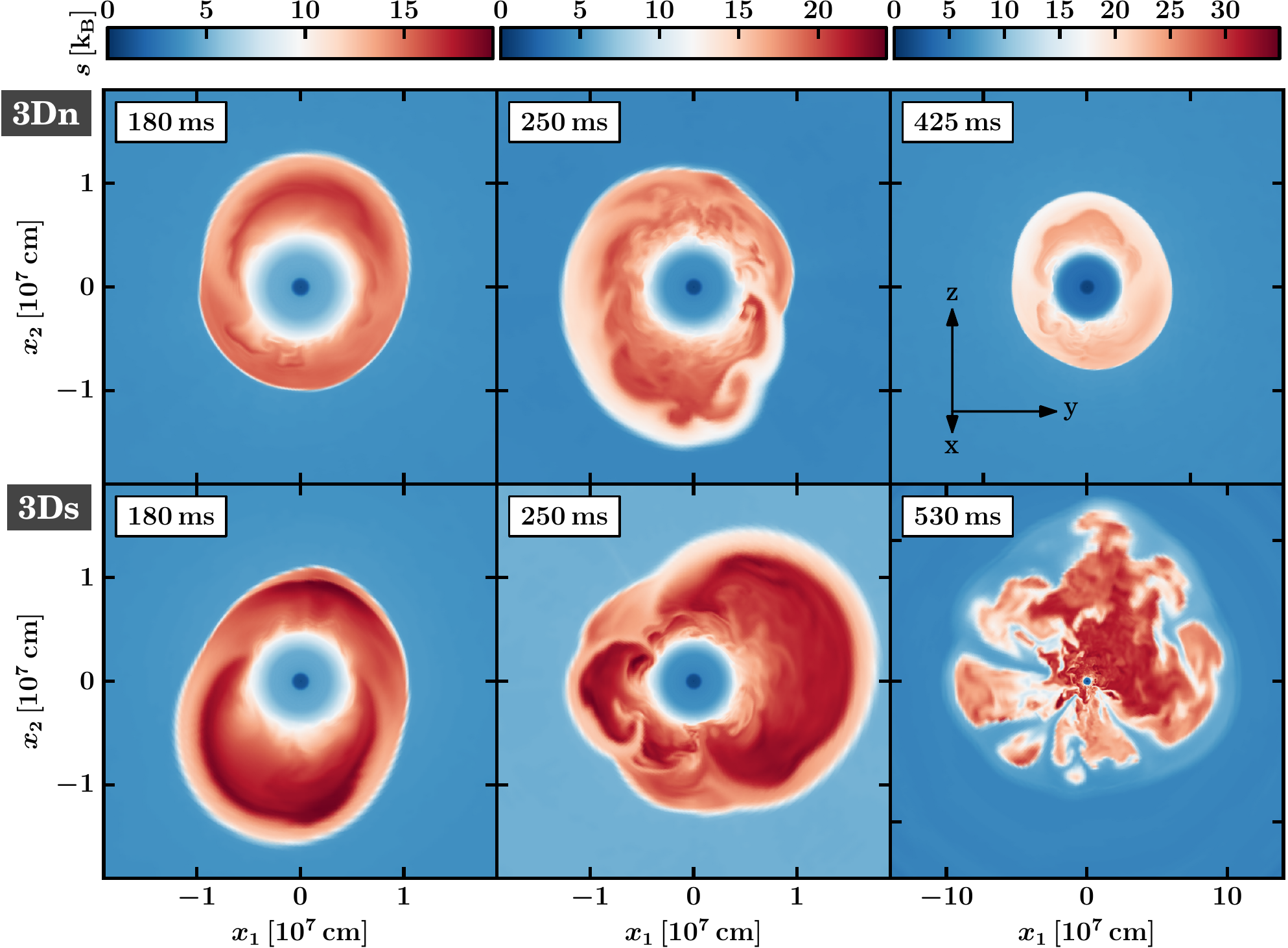}
\caption{
Cross-sectional entropy distributions (in $k_\mathrm{B}$ per nucleon)
for the 3D models without
(3Dn; {\em upper row}) and with strangeness contributions 
(3Ds; {\em bottom}).
The bottom row clearly shows stronger SASI
activity in model 3Ds (180\,ms, 250\,ms), whose traces are still 
imprinted on the ejecta geometry after the onset of
the explosion (530\,ms; note the different scale).
}
\label{fig:3Dn3Ds}
\end{figure*}

While the resolution feasible in current full-scale 3D SN
models is still insufficient to satisfactorily represent postshock
turbulence \citep{abdikamalov14,radice15},
the results reported above suggest that important physics might still 
be missing in the models. One of the aspects to be scrutinized are the
pre-collapse initial conditions, which result from 1D stellar evolution modeling.
\citet{couchott13}, \citet{couch15}, and \citet{mueller15} indeed
confirmed speculations that large-amplitude perturbations
of low-order modes in the convective shell-burning layers 
\citep[e.g.,][and references therein]{arnett11} might 
facilitate the development of explosions. Further progress will require 
3D modeling of the final stages of stellar evolution.

Here we demonstrate that remaining uncertainties in the neutrino 
opacities, in particular the neutrino-nucleon interactions at 
subnuclear densities, can change the 
outcome of 3D core-collapse simulations.
As an example we consider possible strange-quark contributions
to the nucleon spin in their effect on weak neutral-current scatterings.
We show that a moderate isoscalar strange-quark contribution of
$g_\mathrm{a}^\mathrm{s}=-0.2$ to the axial-vector coupling constant
$g_\mathrm{a}=1.26$, which is not far from current experimental results, 
suffices to convert our previous non-exploding 20\,$M_\odot$ 
3D core-collapse run into a successful explosion. 

We briefly describe our numerical approach in Sect.~\ref{sec:numerics},
summarize basic facts about the strange-quark effects in neutrino-nucleon
scattering in Sect.~\ref{sec:strangeness}, discuss our results in
Sect.~\ref{sec:results}, and conclude in Sect.~\ref{sec:conclusions}.

\section{Numerical setup and progenitor model}
\label{sec:numerics}

We performed 2D and full (4$\pi$) 3D simulations of a nonrotating,
solar-metallicity 20\,$M_\odot$ pre-SN progenitor \citep{woosley07}.

We used the \textsc{Prometheus-Vertex} hydrodynamics
code with three-flavor, energy-dependent, ray-by-ray-plus (RbR+) neutrino 
transport including the full set of neutrino reactions and microphysics
\citep{rampp02,buras06} applied in 3D
also by \citet{hanke13} and \citet{tamborra13,tamborra14a,tamborra14b},
in particular the high-density equation of state (EoS) of \cite{lattimer91}
with a nuclear incompressibility of $K=220$\,MeV. To avoid 
time-step constraints, the convectively
stable central core with a radius of 10\,km was computed in spherical
symmetry. We used a 1D gravity potential (which is unproblematic for
strongly PNS-dominated gravity fields) including general relativistic 
corrections \citep{marek06}. The radial
grid had a reflecting boundary condition at the origin and
an inflow condition at the outer boundary of $10^9$\,cm. It had 400
nonequidistant zones initially and was refined in steps up to $>$500
zones, providing an increasingly better resolution of 
$\Delta r/r\,\sim$\,0.01\,...\,0.004
around the gain radius. For the neutrino transport 12 geometrically
spaced energy bins with an upper bound of 380\,MeV were used.
The growth of non-radial hydrodynamic instabilities was seeded by 
imposing random cell-to-cell perturbations of 0.1\% in density on the 
whole computational grid 10\,ms after bounce.

\begin{figure*}
\includegraphics[width=\textwidth]{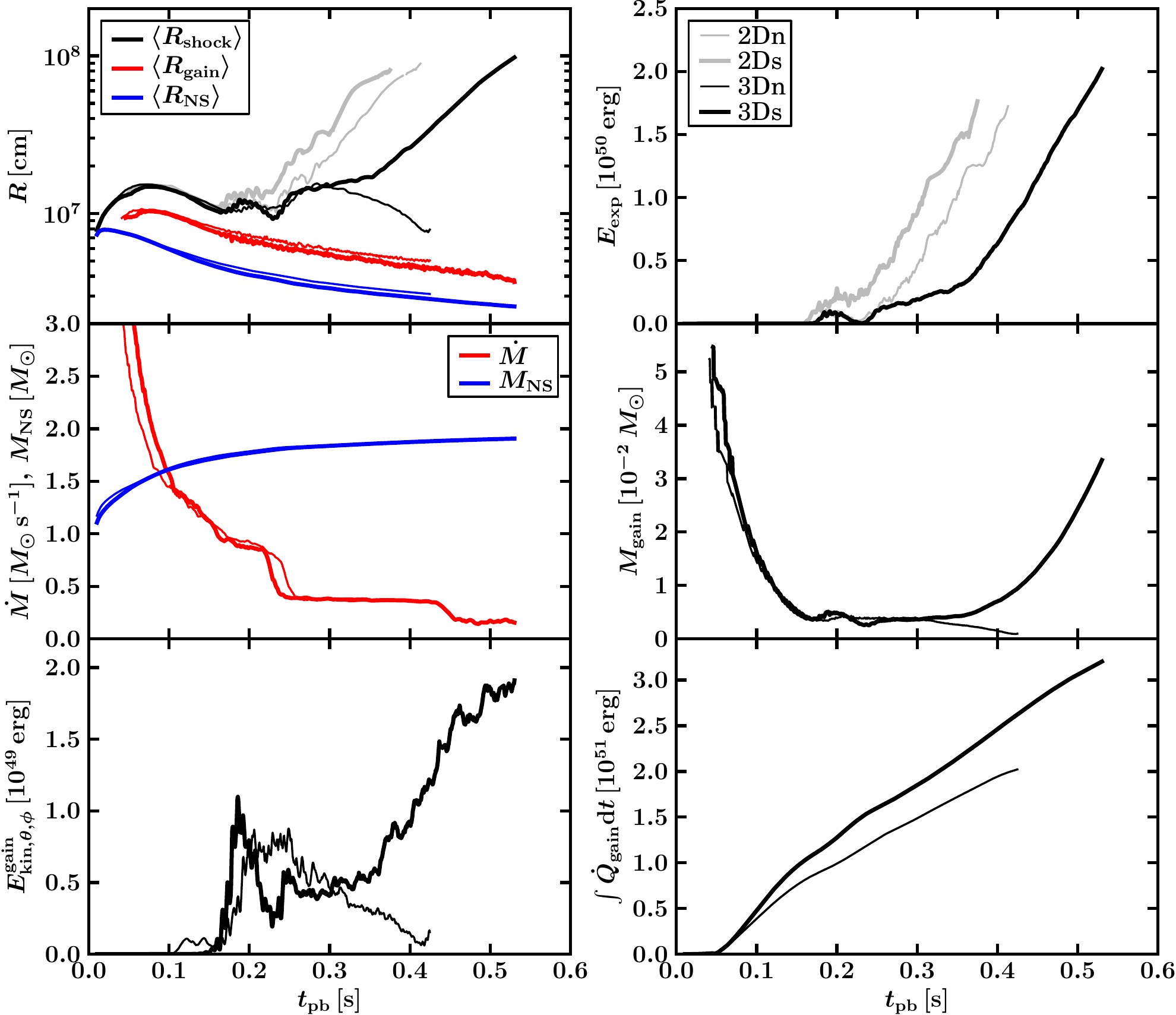}
\caption{
Explosion diagnostics for model 3Ds (thick lines) compared to the
non-exploding model 3Dn (thin lines) as functions of post-bounce
time $t_\mathrm{pb}$.
{\em Top left:} Angle-averaged shock radius (black), gain radius (red)
and NS radius (blue; defined by a density of $10^{11}$\,g\,cm$^{-3}$);
{\em top right:} diagnostic energy (positive total energy behind the
shock). Gray lines display the corresponding 2D models without (2Dn,
thin) and with strangeness contributions (2Ds, thick);
{\em middle left:} mass-accretion rate ($\dot M$) ahead of the shock
(red) and baryonic NS mass (blue);
{\em middle right, bottom left and right:} mass, non-radial kinetic
energy, and time-integrated neutrino-energy deposition in the gain layer,
respectively.
}
\label{fig:explosion}
\end{figure*}

\begin{figure*}
\includegraphics[width=\textwidth]{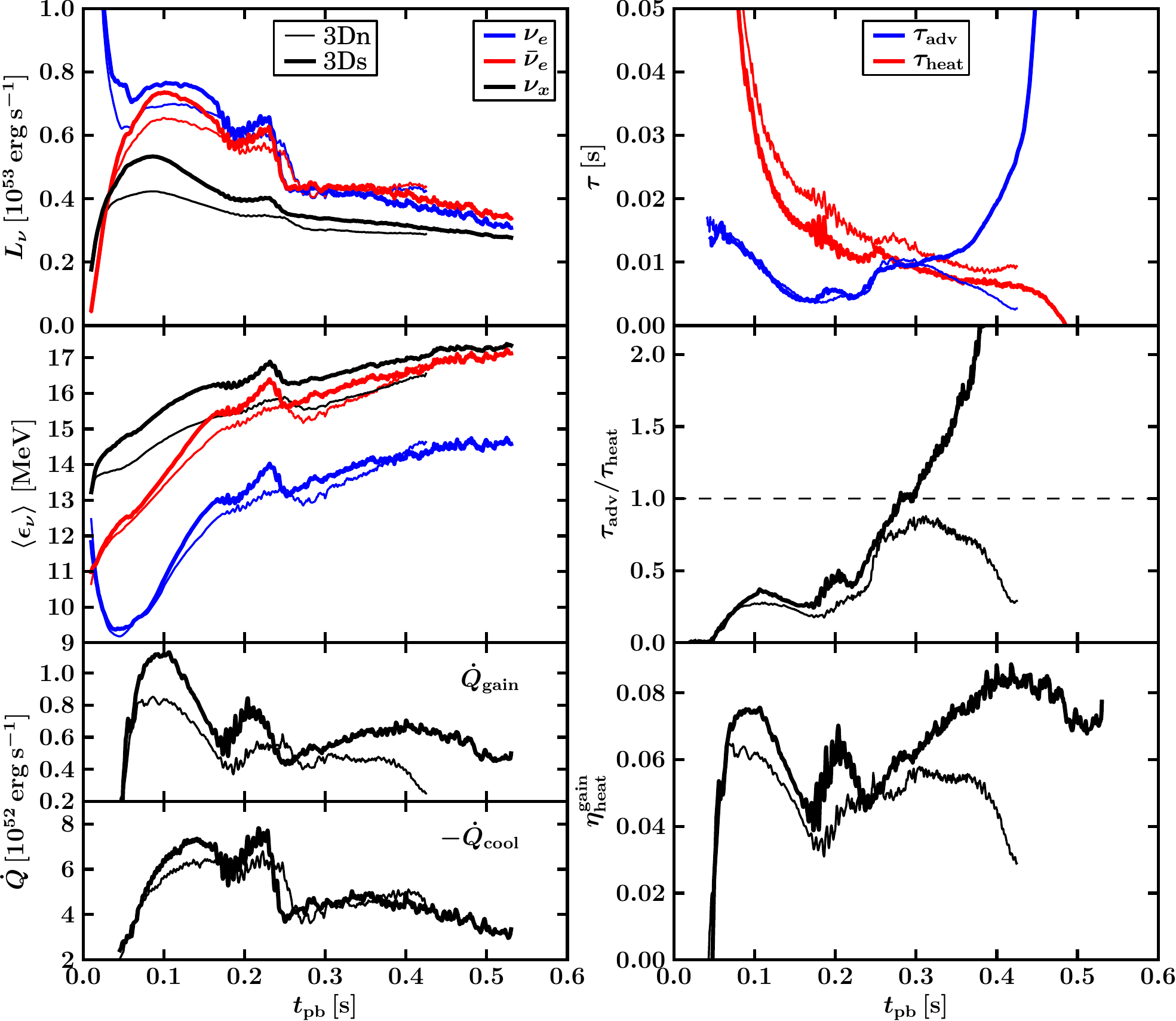}
\caption{
Neutrino-related quantities for model 3Ds (thick lines) compared to
the non-exploding model 3Dn (thin lines) as functions of post-bounce
time $t_\mathrm{pb}$.
{\em Top and middle left:} Neutrino luminosities and mean energies
(defined as ratios of neutrino energy and number fluxes),
respectively, for $\nu_e$ (blue), $\bar\nu_e$ (red), and one kind
of $\nu_x$ (black) for a distant observer at rest;
{\em bottom left:} volume-integrated neutrino-energy deposition rate
in the gain layer (upper panel) and cooling rate between NS radius
and gain radius (lower panel);
{\em top, middle, bottom right:} advection (blue) and heating time scale
(red), corresponding time-scale ratio, and heating efficiency in the
gain layer, respectively.
}
\label{fig:neutrinos}
\end{figure*}

\newpage
\section{Strangeness contributions to neutrino-nucleon scattering}
\label{sec:strangeness}

The lowest-order differential neutrino-nucleon scattering cross section reads
\begin{equation}
\frac{\mathrm{d}\sigma_0}{\mathrm{d}\Omega}=\frac{G_\mathrm{F}^2\epsilon^2}{4\pi^2}
\left [c_\mathrm{v}^2(1+\cos\theta)+c_\mathrm{a}^2(3-\cos\theta)\right ]\,,
\label{eq:diffcross}
\end{equation}
with $\epsilon$ being the incoming neutrino energy, $\theta$ the scattering
angle, $G_\mathrm{F}$ Fermi's constant, and $c_\mathrm{v}$ and $c_\mathrm{a}$
vector and axial-vector coupling constants, respectively. The latter are
$c_\mathrm{v}=\frac{1}{2}-2\sin^2\theta_\mathrm{W}\approx 0.035$, 
$c_\mathrm{a}=g_\mathrm{a}/2\approx0.63$ for ${\nu}p\to{\nu}p$ and
$c_\mathrm{v}=-\frac{1}{2}$, $c_\mathrm{a}=-g_\mathrm{a}/2\approx-0.63$
for ${\nu}n\to{\nu}n$ with $g_\mathrm{a}\approx 1.26$ and 
$\sin^2\theta_\mathrm{W}\approx0.2325$. For iso-energetic scattering
($\epsilon'=\epsilon$), Eq.~(\ref{eq:diffcross}) yields the total
transport cross section
\begin{equation}
\sigma_0^\mathrm{t}=\int_{4\pi}\mathrm{d}\Omega\,\frac{\mathrm{d}\sigma_0}{\mathrm{d}\Omega}(1-\cos\theta)=\frac{2G_\mathrm{F}^2\epsilon^2}{3\pi}\left(c_\mathrm{v}^2+5c_\mathrm{a}^2\right)\,.
\label{eq:totcross}
\end{equation}
While in our SN simulations corrections due to nucleon thermal motions and recoil,
weak magnetism, and nucleon correlations at high densities are taken into account
\citep{rampp02,buras06}, Eqs.~(\ref{eq:diffcross},\ref{eq:totcross}) 
provide good estimates.
Strange quark contributions to the nucleon spin modify $c_\mathrm{a}$ according to
\begin{equation}
c_\mathrm{a}=\frac{1}{2}\left(\pm g_\mathrm{a}-g_\mathrm{a}^\mathrm{s}\right)\,,
\label{eq:gas}
\end{equation}
where the plus sign is for ${\nu}p$ and the minus sign for ${\nu}n$ scattering
\citep[see, e.g.,][]{horowitz02,langanke03}. Since 
$g_\mathrm{a}^\mathrm{s}\le 0$, the cross section for ${\nu}p$-scattering is
increased and for ${\nu}n$-scattering decreased. 

Employing Eq.~(\ref{eq:totcross}) with $g_\mathrm{a}^\mathrm{s}=-0.2$, 
\citet{horowitz02} estimates 15, 21, 23\% reduction of the neutral-current 
opacity for a neutron-proton mixture with electron fractions $Y_e=0.2$, 0.1,
0.05, which are typical values for the layer between neutrinosphere (at density
$\rho\sim 10^{11}$\,g\,cm$^{-3}$) and $\rho\sim 10^{13}$\,g\,cm$^{-3}$ 
for hundreds of milliseconds after bounce.
Since strangeness does not affect charged-current interactions and 
NS matter is neutron-rich, the reduced scattering opacity allows mainly
heavy-lepton neutrinos 
($\nu_x\equiv\nu_\mu,\,\bar\nu_\mu,\,\nu_\tau,\,\bar\nu_\tau$)
to leave the hot accretion mantle of the PNS more easily.
This was found to enhance the expansion of the stalled SN shock in 1D
models, although not enough for successful shock revival 
\citep{liebendoerfer02,langanke03}.
However, below we will show that the situation can be fundamentally
different in 3D simulations.

\begin{figure}
\includegraphics[width=\columnwidth]{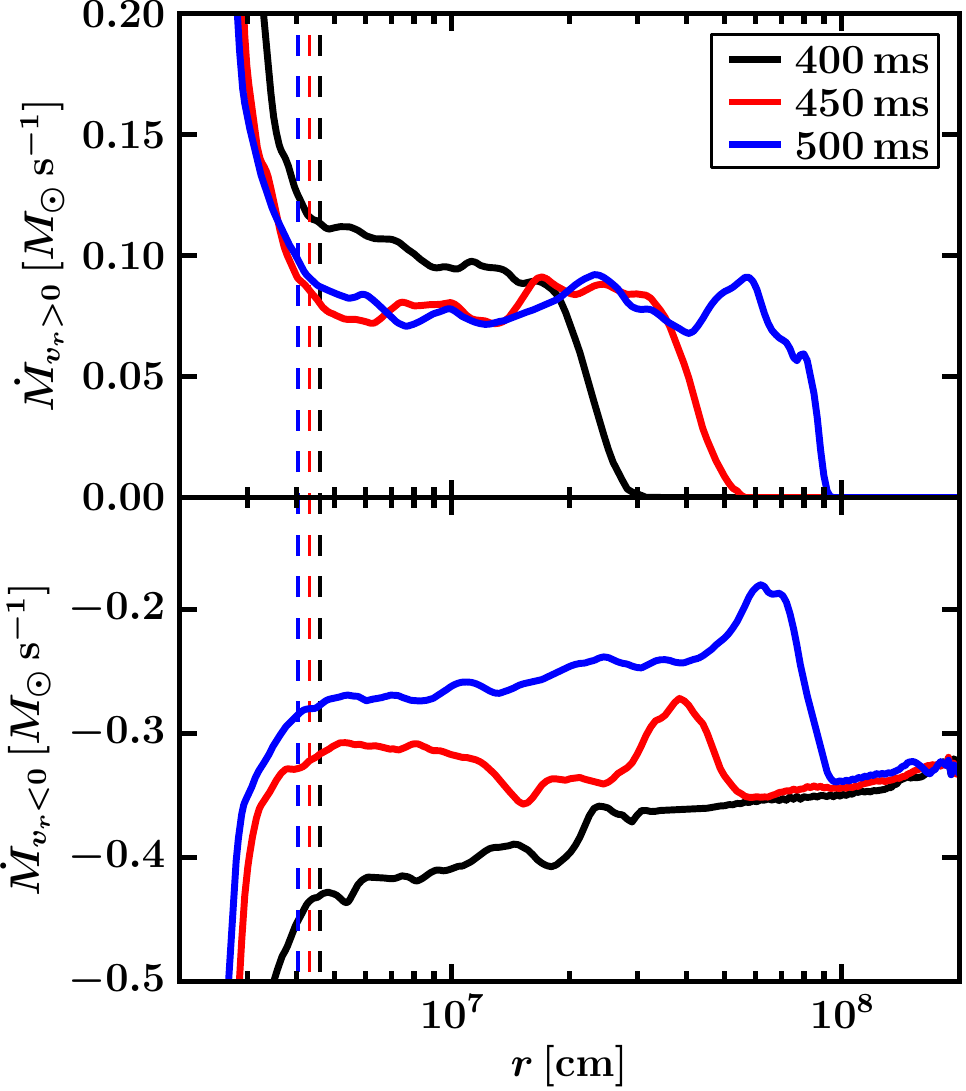}
\caption{
Total mass-accretion rate in inflows ($v_r<0$; {\em bottom}) and mass-outflow
rate in rising plumes in the gain layer ($v_r>0$; {\em top}) at three times
after the onset of explosion of model 3Ds. Vertical dashed lines mark
the positions of the average gain radius, the steep forward edges indicate
the average shock radii.
}
\label{fig:accretion}
\end{figure}

\section{Results}
\label{sec:results}

We compare 2D and 3D core-collapse simulations of the 20\,$M_\odot$ star
with strangeness corrections in neutrino-nucleon scatterings, using
$g_\mathrm{a}^\mathrm{s}=-0.2$ (models 2Ds, 3Ds), to corresponding 
simulations without strange quark effects ($g_\mathrm{a}^\mathrm{s}=0$;
models 2Dn, 3Dn) as in all SN simulations of the Garching group so far.
To explore ``extreme'' effects, our choice of 
$g_\mathrm{a}^\mathrm{s}$ is by its absolute value somewhat bigger than
theoretical and experimental determinations of
$g_\mathrm{a}^\mathrm{s}\sim -0.1$ \citep{ellis97,alexakhin07,airapetian07}.

\subsection{Dynamical evolution towards explosion}

The sequence of 3D images in Fig.~\ref{fig:3Dsevol} shows the post-bounce
evolution of the exploding model 3Ds; the entropy cuts in Fig.~\ref{fig:3Dn3Ds}  
demonstrate the differences to the unsuccessful model 3Dn. In both cases
the dynamics of the accretion layer are strongly SASI-dominated. In model
3Ds there is no indication of postshock convection before SASI sloshing
motions first appear at post-bounce (p.b.) time $t_\mathrm{pb}\sim 120$\,ms.
These reach full strength around 180\,ms p.b.\ and continue in varying
directions until $\sim$280\,ms. Only later on convective overturn takes
over as the dominant non-radial instability. In model 3Dn moderate buoyancy
activity is visible from 100--180\,ms before SASI becomes dominant, too.
Early convection is enabled by a slightly larger radius of the stalled 
shock in 3Dn as a consequence of a slightly lower mass-accretion rate $\dot{M}$
(Fig.~\ref{fig:explosion}). This difference improves the growth conditions for 
postshock convection. In 3Dn an erroneous change of the transition
between low-density and high-density EoS caused a delay of the core collapse.
Therefore the mass-accretion rate until $t\sim 150$\,ms is slightly reduced
and the Si/Si+O interface arrives at the shock $\sim$15\,ms later. These
early differences are inessential for our discussion because neutrino heating 
creates favorable conditions for the explosion of model 3Ds only later than
$\sim$300\,ms.

From $t_\mathrm{pb}\sim 170$\,ms on, 3Ds exhibits clearly larger SASI
amplitudes and higher postshock entropies, which increases the maximum and 
average shock radii (Figs.~\ref{fig:3Dn3Ds}, \ref{fig:explosion}). This
model also shows larger non-radial kinetic energies in the gain layer,
\begin{equation}
E_{\mathrm{kin},\theta,\phi}^\mathrm{gain}=\int_{\langle R_\mathrm{gain}\rangle}^{
R_\mathrm{shock}(\theta,\phi)}\ud V\,\frac{1}{2}\rho\left(v_\theta^2+v_\phi^2\right)
\end{equation}
(Fig.~\ref{fig:explosion}), except during $t_\mathrm{pb}\sim200$--300\,ms, when
a powerful spiral SASI mode develops in model 3Dn but not in 3Ds, albeit
without pushing the shock sufficiently far out for revival (in conflict
with recent results by \citealp{fernandez15}).

At $t_\mathrm{pb}\gtrsim 300$\,ms, roughly 50\,ms after the Si/Si+O interface
has fallen through the shock and the shock has expanded to $\sim$150\,km,
the evolutions of models 3Dn and 3Ds separate.
While in 3Dn the average shock radius, $\left< R_\mathrm{shock}\right>$, 
retreats again and conditions become unfavorable for an explosion, model
3Ds exhibits positive trends in all explosion-diagnostic parameters 
(e.g., $\left<R_\mathrm{shock}\right>$,
$E_{\mathrm{kin},\theta,\phi}^\mathrm{gain}$,
$M_\mathrm{gain}$). Continuous shock expansion signals runaway and finally 
outward acceleration sets in at $t_\mathrm{pb}\gtrsim 360$\,ms, at which time 
the recombination of free nucleons to $\alpha$-particles in the largest plumes
begins to release energy and the instantaneous ``diagnostic energy'',
\begin{equation}
E_\mathrm{exp}=\int_{e_\mathrm{tot}>0,\mathrm{postshock}}\!\ud V\,\,\rho
e_\mathrm{tot}\,,
\label{eq:expenergy}
\end{equation}
starts to rise steeply, correlated with a fast growth of the mass in the gain
layer (Fig.~\ref{fig:explosion}). In Eq.~(\ref{eq:expenergy}) the volume 
integration is performed over the postshock region where the total specific 
energy,
\begin{equation}
e_\mathrm{tot}=e+\frac{1}{2}|\textit{\textbf{v}}|^2+\Phi+\left[
e_\mathrm{bind}(^{56}\mathrm{Fe})-e_\mathrm{bind}\right]\,,
\label{eq:totenergy}
\end{equation}
is positive, with $e$, $\frac{1}{2}|\textit{\textbf{v}}|^2$, and $\Phi$
being the specific internal, kinetic, and (Newtonian) gravitational energies.
The bracketed term expresses the difference between the specific nuclear binding
energy when all nucleons are finally recombined to iron-group nuclei
compared to the nuclear composition at a given time. It therefore accounts for
the maximum release of nuclear energy and corresponds to an upper limit
of $E_\mathrm{exp}$. 

Both corresponding 2D models, 2Dn and 2Ds, also explode after the 
Si/Si+O interface has passed the shock, but
in 2Ds the outward shock acceleration and rise of $E_\mathrm{exp}$ sets in
$\sim$50\,ms earlier (Fig.~\ref{fig:explosion}). Strangeness corrections
therefore create more favorable explosion conditions also in the 2D case,
although their influence is modest in successful models, similar to their
small effect in the 1D case, which is far away from explosion 
\citep{liebendoerfer02}.

\subsection{Strangeness corrections and explosion}

The impact of strangeness effects on the neutrino emission and the
explosion is displayed in Fig.~\ref{fig:neutrinos}. Model 3Ds consistently
exhibits higher luminosities and mean energies of the emission for
all neutrino species, and, consequently, a higher neutrino-energy
deposition rate in the gain layer, $\dot{Q}_\mathrm{gain}$, a higher
heating efficiency,
\begin{equation}
\eta_\mathrm{heat}^\mathrm{gain}=\frac{\dot{Q}_\mathrm{gain}}{L_{\nu_e}+L_{\bar{\nu}_e}}\,,
\end{equation}
a smaller gain radius (Fig.~\ref{fig:explosion}),
and a shorter heating time scale,
\begin{equation}
\tau_\mathrm{heat}=\frac{\left|E_\mathrm{gain}\right|}{\dot{Q}_\mathrm{gain}}
\end{equation}
with $E_\mathrm{gain}=
\int_{\langle R_\mathrm{gain}\rangle}^{R_\mathrm{shock}(\theta,\phi)}\ud V
\rho(e+\frac{1}{2}|\textit{\textbf{v}}|^2+\Phi)$ being the binding energy of 
the gain layer. Since the effective time scale of mass advection through the 
gain layer,
\begin{equation}
\tau_\mathrm{adv}=\frac{M_\mathrm{gain}}{\dot{M}}
\end{equation}
(where $\dot{M}>0$), which measures the average exposure time of matter to 
neutrino heating, is very similar in models 3Ds and 3Dn, the smaller 
$\tau_\mathrm{heat}$ in 3Ds also leads to a higher time-scale ratio
$\tau_\mathrm{adv}/\tau_\mathrm{heat}$. The ratio 
$\tau_\mathrm{adv}/\tau_\mathrm{heat}$ exceeds the critical
value of unity shortly before the SN shock in 3Ds begins its runaway expansion.

The mean energies of the radiated neutrinos in model 3Ds are up to $\sim$1\,MeV 
higher and the luminosities of $\nu_e$ and $\bar\nu_e$
by up to $\sim$10--15\%, whereas the $\nu_x$-luminosities rise by up to $\sim$30\%.
The increase of the total neutrino luminosity is more than 
$6\times 10^{52}$\,erg\,s$^{-1}$
at maximum, which mainly comes from layers below the $\nu_e$-sphere, because
the neutrino-loss rate $\dot Q_\mathrm{cool}$ between the location of this sphere
(at $\sim$10$^{11}$\,g\,cm$^{-1}$) and the gain radius differs between models
3Ds and 3Dn by at most $\sim$10$^{52}$\,erg\,s$^{-1}$ (Fig.~\ref{fig:neutrinos}).
Note that at $t_\mathrm{pb}\gtrsim 300$\,ms the relative differences of the
neutrino properties of models 3Ds and 3Dn decrease and even change sign, 
because the former explodes whereas the latter continues to collapse and to
accrete mass onto the PNS at a higher rate.

For $Y_e=0.1$--0.05, strangeness effects in the neutrino-nucleon interactions
reduce the effective opacity
$\kappa_\mathrm{eff}=\sqrt{\kappa_\mathrm{abs}(\kappa_\mathrm{abs}+\kappa_\mathrm{scatt})}$ 
for $\nu_e$ only by 2--3\% and for $\bar\nu_e$ by 8--10\%. 
A considerable part of the observed 
luminosity enhancement of $\nu_e$ and $\bar\nu_e$ is therefore caused
indirectly by a stronger contraction of the PNS in response to the larger 
energy loss through $\nu_x$ emission. 
With the smaller PNS radius (Fig.~\ref{fig:explosion}) and steeper density 
profile, the neutrinospheres of $\nu_e$ and $\bar\nu_e$ move inward to higher
temperatures.

The reduction of the weak neutral-current scattering by strange quark 
contributions to the nucleon spin therefore enhances the neutrino luminosities
and mean energies directly and indirectly and thus strengthens the neutrino
heating in the gain layer. This amplifies buoyancy and turbulent mass motions
behind the shock, which is signaled by higher non-radial kinetic energy in
model 3Ds at $t_\mathrm{pb}\gtrsim 300$\,ms (Fig.~\ref{fig:explosion}), thus
fostering the explosion of this model in contrast to 3Dn.

\subsection{Further development of explosion energy}

At the end of the simulation at 530\,ms p.b., the PNS has a baryonic
mass of 1.91\,$M_\odot$ and the shock has expanded to an average radius of
$\left<R_\mathrm{shock}\right>\approx 1000\,$km. The diagnostic energy,
$E_\mathrm{exp}$, has reached $0.2\times 10^{51}$\,erg and rises steeply and
linearly with a rate of
$\dot{E}_\mathrm{exp}\approx 1.2\times 10^{51}$\,erg\,s$^{-1}$.
This growth rate can be understood by the ejection of a continuous outflow of
freshly neutrino-heated matter, $\dot{M}_{v_r>0}\sim 0.08$--0.1\,$M_\odot$\,s$^{-1}$
(Fig.~\ref{fig:accretion}; the subscript indicates positive radial velocity),
which is mainly fed by the shock-accreted matter that is channeled to the
gain radius in persistent accretion downdrafts (Fig.~\ref{fig:3Dn3Ds}) with
mass-inflow rates up to $\sim$0.3\,$M_\odot$\,s$^{-1}$ (Fig.~\ref{fig:accretion}).
The mass outflow does not only absorb neutrino energy, it also releases
nuclear binding energy from the recombination of neutrons and protons to
$\alpha$-particles and heavy nuclei. Since neutrino heating roughly neutralizes
the gravitational binding energy of the matter \citep{janka01,marek09},
the growth of $E_\mathrm{exp}$ can be estimated in terms of 
$\dot{M}_{v_r>0}$ and an average nuclear recombination energy per nucleon, 
$\epsilon_\mathrm{nuc}$, as
\begin{equation}
\dot{E}_\mathrm{exp}\approx\dot{M}_{v_r>0}\,\frac{\epsilon_\mathrm{nuc}}{m_\mathrm{B}}
\approx 1.2\times 10^{51}\,\left(\frac{\dot M_{v_r>0}}{0.1\,M_\odot/\mathrm{s}}\right)
\left(\frac{\epsilon_\mathrm{nuc}}{6\,\mathrm{MeV}}\right)\,
\frac{\mathrm{erg}}{\mathrm{s}}\,.
\label{eq:dotexp}
\end{equation}
To unbind the overlying layers of the progenitor with a gravitational binding
energy of $\sim$0.6$\times{10}^{51}$\,erg would therefore require massive outflow
for several 100\,ms. The neutrino-driven wind after the end of accretion as well
as explosive nuclear burning will provide additional energy. For a reliable
determination of the final SN explosion energy, model 3Ds would need to be evolved 
considerably longer than in our simulation.

\section{Conclusions}
\label{sec:conclusions}

We showed that strangeness contributions to neutrino-nucleon
scattering with an axial-vector coupling of $g_\mathrm{a}^\mathrm{s}=-0.2$
are sufficient to turn a non-exploding 3D simulation of a 20\,$M_\odot$ model 
(in which $g_\mathrm{a}=1.26$ was used for the standard isovector form factor)
to a successful explosion. Strange-quark effects in the nucleon spin reduce the 
neutrino opacity of neutron-rich matter inside the neutrinosphere and thus
directly and indirectly enhance the luminosities and mean energies of the radiated
neutrinos. This leads to amplification of the neutrino-energy deposition behind
the stalled shock because charged-current interactions are not affected by nucleon
strangeness. Owing to a reduced neutrino-heating time scale and stronger non-radial
mass motions in the gain layer, 
the shock is driven to runaway expansion $\sim$100\,ms
after the passage of the Si/Si+O interface. The enhanced neutrino emission enabled
by the strangeness effects is associated with (and partly caused by) a faster 
contraction of the PNS and a corresponding rise of the neutrinospheric
temperatures. Strangeness corrections in neutrino-nucleon scattering are therefore 
similarly beneficial for shock revival as ``softer'' nuclear EoSs, which also
lead to faster PNS contraction and stronger emission of more energetic neutrinos
\citep{marek09,janka12,suwa13}.

Our results demonstrate how close previous, unsuccessful 3D 
core-collapse models with the \textsc{Prometheus-Vertex} code were to explosion. 
They also underline that an accurate knowledge of neutrino-nucleon 
interaction rates, in particular also for neutral-current scattering, 
is of crucial importance for assessing the viability of the neutrino-driven
explosion mechanism. While strangeness contributions affect
neutrino-nucleon scattering everywhere, such opacity modifications
between the subnuclear regime and neutrinospheric densities are 
most relevant for SN shock revival. 
In the discussed 20\,$M_\odot$ simulations a modest $\sim$15\% 
diminution of neutrino-neutron scattering makes all the difference between
explosion and failure. 
Theoretical and experimental determinations yield
somewhat smaller absolute values for $g_\mathrm{a}^\mathrm{s}$ than
assumed in our study \citep{ellis97,alexakhin07,airapetian07}.
However, in-medium effects like correlations in low-density nucleon
matter may cause similar opacity reductions (C.~Horowitz, private 
communication).

Nucleon-strangeness effects should also be investigated in
3D simulations of other progenitors. Moreover, our calculations must be
repeated with better than 2$^\circ$ angular zoning to ensure that the
explosion is robust and withstands higher
angular resolution of the cascading of turbulent energy from the
largest scales to small structures, which can be harmful for
shock revival \citep{hanke12,couch13,couch14,abdikamalov14}.
In any case, however, the outcome of multi-dimensional core-collapse 
simulations that marginally explode or fail can sensitively
depend on effects on the 10\% level in the neutral-current
neutrino-nucleon interactions. This sensitivity needs to be taken into
account in numerical implementations of these rates and might
also be important for understanding partially conflicting model results
published by different groups.

\acknowledgements
We thank C.~Horowitz, G.~Raffelt and S.~Wanajo for comments
and Elena Erastova (RZG) and Aaron D\"oring for visualization support.
At Garching, the project was funded by grant EXC~153 from DFG,
ERC-AdG No.~341157-COCO2CASA, and computing time from the European
PRACE Initiative on SuperMUC (GCS@LRZ, Germany) and MareNostrum 
(BSC, Spain). B.M.\ acknowledges support by ARC
through DECRA grant DE150101145.

\bibliographystyle{apj}

\end{document}